\documentclass{sigchi}
\usepackage{float}

\usepackage{balance}       
\usepackage{graphics}      
\usepackage[T1]{fontenc}   
\usepackage{txfonts}
\usepackage{mathptmx}
\usepackage[pdflang={en-US},pdftex]{hyperref}
\usepackage{color}
\usepackage{booktabs}
\usepackage{textcomp}
\usepackage{graphicx}

\usepackage{microtype}        
\usepackage{ccicons}          

\usepackage{todonotes}

\def\plaintitle{PARS3: Parallel Sparse Skew-Symmetric Matrix-Vector Multiplication with Reverse Cuthill-McKee Reordering  }

\def\emptyauthor{}
\def\plainkeywords{symmetric; sparse;  matrix vector multiplication; reverse cuthill mckee; reordering; matrix bandwith;}

\makeatletter
\def\url@leostyle{%
  \@ifundefined{selectfont}{
    \def\UrlFont{\sf}
  }{
    \def\UrlFont{\small\bf\ttfamily}
  }}
\makeatother
\def\pprw{8.5in}
\def\pprh{11in}

\definecolor{linkColor}{RGB}{6,125,233}
\hypersetup{%
  pdftitle={\plaintitle},
  pdfauthor={\emptyauthor},
  pdfkeywords={\plainkeywords},
  pdfdisplaydoctitle=true, 
  bookmarksnumbered,
  pdfstartview={FitH},
  colorlinks,
  citecolor=black,
  filecolor=black,
  linkcolor=black,
  urlcolor=linkColor,
  breaklinks=true,
  hypertexnames=false
}

\makeatletter
\def\@copyrightspace{\relax}
\makeatother

\begin{document}

\title{\plaintitle}

\numberofauthors{3}
\author{%
  \alignauthor{Selin Yıldırım\\ 
    \affaddr{Department of Computer Science, University of Illinois Urbana-Champaign}\\
    \affaddr{Illinois, United States}\\
    \email{seliny2@illinois.edu}}\\
  \alignauthor{Murat Manguoğlu\\
    \affaddr{Department of Computer Engineering, Middle East Technical University}\\
    \affaddr{Ankara, Turkey}\\
    \email{manguoglu@ceng.metu.edu.tr}}\\
}

\maketitle

\begin{abstract}

  Sparse matrices, as prevalent primitive of various scientific computing algorithms,  persist as a bottleneck in processing. A skew-symmetric matrix flips signs of symmetric pairs in a symmetric matrix. Our work, Parallel 3-Way Banded Skew-Symmetric Sparse Matrix-Vector Multiplication, equally improves parallel symmetric SpMV kernels with a different perspective than the common literature trends, by manipulating the form of matrix in a preprocessing step to accelerate the repeated computations of iterative solvers. We effectively use Reverse Cuthill-McKee (RCM) reordering algorithm to transform a sparse skew-symmetrix matrix into a band matrix, then efficiently parallelize it by splitting the band structure into 3 different parts by considering its local sparsity. Our proposed method with RCM is novel in the sense that it is the first implementation of parallel skew-symmetric SpMV kernels.  Our enhancements in SpMV and findings are valuable with significant strong scalings of up to $19\times$ over the serial compressed SpMV implementation. We overperform a heuristic-based graph-coloring approach with synchronization phases in implementing parallel symmetric SpMVs. Our approach also naturally applies to parallel sparse symmetric SpMVs, that can inspire widespread SpMV solutions to adapt presented optimizations in this paper.
\end{abstract}
\keywords{\plainkeywords}

\section{1. Introduction}
 
Scientific computing applications quite often involve numerical algorithms, whose  primitives are matrix (or, equivalently graph) operations. Common sparsity observed in matrices challenge software performance, requiring performance tuning at lower levels, such as tuning the solution for  memory hierarchy, bandwith and computing resources at specific hardware. Sparse matrices, as irregular data structures, when parallelized not cautiously, exhibit poorly organized memory access patterns that creates a major bottleneck.

Sparse Matrix-Vector multiplication (SpMV) is such widely used computational kernel in applications of artificial intelligence, data science, physical simulations, statistical analysis and many others. One closely related example is the solution of linear system of equations. Depending on the numerical and structural properties of the coefficient matrix, various classes of iterative or direct linear system solvers have been developed. Iterative solvers require a matrix vector multiplication with the coefficient matrix in each iteration. 

Given a sparse skew-symmetric linear system of equations, 
\begin{equation}
Ax = b 
\end{equation}
where $A\in\mathbb{R}^{n\times n}$ and $x,b\in\mathbb{R}^{n}$ if the coefficient matrix, $A$,  is skew-symmetric (i.e. $A=-A^T$). In some applications, $A$ is naturally skew-symmetric, while in many others it is "near" skew-symmetric with $A=\alpha I + S$ (also known as a shifted skew-symmetric matrix) where $\alpha$ is a scalar and $S$ is skew-symmetric or in the form $A=J-S$ where $J$ is symmetric positive semi-definite. Again, the latter can be transformed into the shifted skew-symmetric form easily
\cite{guducu2022non,iterativeSolver}
. Furthermore,  when the coefficient matrix is just general (i.e. virtually in every application), with some work one can still precondition the matrix to to obtain a near skew-symmetric form \cite{iterativeSolver}.

There are many iterative schemes proposed for solving skew- and shifted skew-symmetric systems in the literature,  most notably Krylov subspace based  MRS method\cite{jiang2007algorithm,idema2007minimal}. The striking feature of MRS iterative scheme is that it only requires one matrix-vector multiplication and one inner-product operation per iteration. Inner products create synchronization points which severely limit the overall parallel scalability of the iterative scheme which can potentially degrade the parallel scalability significantly. While Conjugate Gradient (CG) algorithm also requires the same amount of operations, CG requires the coefficient matrix to be symmetric and positive definite, therefore is more restrictive and there is no other algorithm with these features that are applicable to general problems. 

Skew-symmetric or nearly skew-symmetric systems arise in  Navier-Stokes Equations \cite{exahype} for simulating fluid flow, Linear Least Squares and optimization problems, such as linear regression. Quantum Mechanics also involves Hermitian Matrices which are generalization of skew-symmetric matrices. 

Not only inner products, but sparse matrix vector multiplications are typically costly in iterative solvers.  After compact representations for sparse matrix storage schemes emerged, parallel solutions to SpMV methods also appeared. Compared to the dense counterparts, sparse matrix multiplication bring extra challenges to memory, and communication costs.  
Two commonly used storage schemes, namely Symmetric Sparse Skyline (SSS) and Compressed Row Storage (CRS), rely on an array of row pointers and their column based counterparts. These row pointers indicate the nonzero counts in each compressed row, along with their corresponding original column indices and values. SSS differs from CRS by storing nonzero matrix elements located at either lower or upper triangular of the symmetric matrices, whereas CRS stores all nonzeros of any matrix which is not necessarily symmetric. Being slightly different than CRS and SSS formats, Coordinate (COO) format stores row and column indices, as well as numerical value of each nonzero matrix element. 

Race conditions appear in many parallel programs along with its synchronization costs. In multi-threaded and multi-processed applications, a data race refers to reading a different value from a memory location depending on the CPU order of multiple execution units' update on the same, shared location in memory. In other words, the memory reads a value based on the relative order of process/thread updates on the same location. 

 Data races lead to unpredictable behavior as the order in which different threads or processes access shared data is not guaranteed, which may also result in bugs, performance and security issues [9]. To mitigate these issues, proper synchronization mechanism such as mutexes and locks, semaphores, atomic operations, barriers, memory fences and thread-safe data structures might be used to ensure program correctness. However, introducing synchronization primitives into a program comes with its own costs. Critical sections allow only one thread to execute at a time for the given code block, which prevents the remaining threads from proceeding with execution as they are required to wait on the same synchronization primitive. Spinning on a synchronization variable by some threads is such a scenario where we observe serialization and overall performance degradation. In the nature of sparse matrix vector multiplication, we observe these challenges during accesses to shared output locations by different processes, which requires synchronization to obtain correct results.

In graph-coloring approach to parallel SpMV\cite{Athens}, the number of data races with the benchmark matrices are given.  The number of data races tend to increase with the number of processes. Hence, the overhead of parallelism in SSpMV, which mainly originates in shared output data among processes, is also increasing. The more execution units are introduced into solving the problem, the more overlap on the identical memory locations happen during parallel processing.
In \cite{Athens}, the authors investigate possible independent sets of elements that are safe to run in parallel, by eliminating as many dependent sets of nonzeros as possible that would cause races in the same phase. Authors apply graph coloring and row decomposition methods to obtain more independent sets in graphs. Exploring independent sets of data to be processed concurrently is one of the trends in the literature for fast processing symmetric SpMV kernels. However, common disadvantages come with the existing synchronization points between dependent phases, by jeopardising parallelism and spoiling the memory access patterns, especially when number of phases are considerably large. In addition, it was pointed out that high-bandwidth matrices are more likely to perform poorly due to the fact that they contain more race elements and yield much less independent nonzero sets. Authors indicate grouping different nonzeros of a matrix, also known as separating matrix bandwiths, and processing them at different stages may be promising to reduce race conditions. We emphasize at this point that, in addition to matrix splitting based on amount of sparsity, matrix reordering algorithms are also helpful to reduce matrix bandwidth by transforming the original sparse matrix into a band form.  Towards Fast SpMV\cite{ETH}  extensively discusses potential power of the reordering method and demonstrates how significantly it can contribute to the overall performance of a SpMV kernel. 

\begin{figure}[H]
\centering
  \includegraphics[width=0.9\columnwidth]{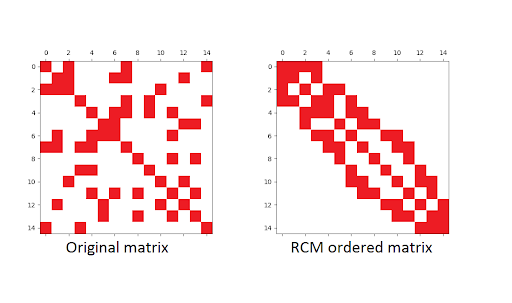}
  \caption[]{Demonstration of RCM algorithm \footnotemark }
  \label{fig:figure1}
\end{figure}

With our approach, we apply the promising reordering method to focus on improving the kernel by transforming its general pattern into a more structured data, the band form, in Figure \ref{fig:figure1}, so that the sparsity characteristics of the several spatial regions over the band can be exploited under multicore algorithms. In the preprocessing stage, we first apply RCM by using MATLAB and split the band matrix into 3 parts with respect to their sparsity. As illustrated in Figure \ref{fig:figure7} and \ref{fig:figure8}, the first split constitutes for the main diagonal of the band matrix, which is conventionally dense in a band matrix. Remaining elements in lower and upper triangular part of the band matrix are the elements composing the band form at the center (in the middle) of the horizontal band, and outer elements are clustered over the margins of the band. We determine and group these two splits, middle and outer splits, according to the particular bandwidth given as a user input. Middle split is observed to exhibit high sparsity compared to other splits due to the RCM transformation. With the same reason, number of elements in the outer split is expected to be dramatically less, as opposed to the split extracted from the middle.

To summarize, the contribution of our paper is :

- Fast 3-way parallel implementation of skew-symmetric sparse matrix vector multiplication by using RCM reordering with SSS storage and MPI, in $\mathbf{\theta}$(NNZ) time. We show our solution analyzes Amdahl's speedup and improves this metric over the conflict-free SSpMV\cite{Athens} that uses synchronization phases. 
\footnotetext{Image credit to www.juliafem.org}

\begin{table}[H]
\scalebox{0.95}{
\begin{tabular}{| l | r | r | r |}
\hline
\textbf{Matrix } & \textbf{$\#$ Rows}   & \textbf{$\#$ Nonzeros}  &  \textbf{RCM Bandwith}     \\ \hline
\textbf{boneS10} & 914,898   & 40,878,708 & 13,727 \\ \hline
\textbf{Emilia$\_$923}  & 923,136   & 40,373,538 & 14,672 \\ \hline
 \textbf{ldoor} & 952,203   & 42,493,817 & 8,707  \\ \hline
\textbf{af$\_$5$\_$k101} & 503,625   & 17,550,675 & 1,274  \\ \hline
 \textbf{Serena} & 1,391,349 & 64,131,971 & 87,872 \\ \hline
\textbf{audikw$\_$1}  & 943,695   & 77,651,847 & 35,102 \\
\hline
\end{tabular}
}
\vspace{0.75cm}
\caption{Characteristics of our benchmark matrices\protect\cite{SPARSKIT} }~\label{tab:tab1}
\end{table}

\section{2. Related Work}
As parallel computing becomes more prominent in scientific computing domains, researchers implement many variants of multicore SSpMVs by using numerous techniques and efficient matrix representations to process big data. Most of the problems involving sparse matrices and graphs have remained as a bottleneck in numerical computations.

A comprehensive introduction to computing with irregular data structures and their complications is discussed in the paper \cite{irregular}. Authors develop a hardware neutral profiling tool, Parameter, as an abstract measure of the amount of amorphous data-parallelism at run-time. Amorphous data-parallelism exists where some set of concurrent operations in a parallel program conflict with each other. Amorphous data structures such as graphs and sparse matrices put another limit on parallelism due to their compute size changing dynamically as required by some algorithms. On the other hand, fixed size problems such as matrix-matrix multiplication, and SpMVs do not exhibit the growing work volume and have a  constant number of essential operations for a constant input data size. On the contrary, for some other algorithm classes (e.g., Delaunay Mesh Refinement\cite{irregular} ), decisions made at runtime on dynamic operations, in other words conflicting operations, may result in different number of total operations within the concept of amorphous data-parallelism. Authors prefer to resolve such  conflicts by incurring some overhead of involving optimistic parallelism into their solution, in Galois system with speculative execution support to rollback the execution upon conflicts. However, this approach may not be always useful for the systems where optimistic parallelism is not preferrable.

A sequential algorithm as a baseline is given in \cite{Athens}. Figure \ref{fig:figure3} provides the pseudocode of the algorithm which uses SSS format. The algorithm has time complexity of $\mathbf{\theta}$(NNZ) and is memory-bound. 


Basic Linear Algebra Subprograms (BLAS) library contains a general double presicion banded matrix vector multiplication (dgbmv) algorithm  which performs the matrix vector multiplication after compressing the nonzero elements over band form into LAPACK banded storage format, in order to achive better memory utilization. The disadvantage of using LAPACK banded storage is however the wasted storage in rectangular shaped arrays due to the zeros around the band.

We are inspired by the experiments conducted in \cite{ETH}  that evaluates how crucial matrix reordering is for accelerating SpMVs on CPUs. Authors implement Cuthill-McKee (CM) reordering to show that it reduces cache misses incurred by accessing
to $x$ (input) and $y$ (output) vectors, as well as decreasing the communication volume among parallel processes.  The paper implements 3 variants of Symmetric SpMV, which are with and without latency hiding that impose particular orders on messages in a way that exchanging vector portions overlaps with partial multiplications. First and second routine does not scale well and they are dedicated to machines with no support for latency-hiding. The other drawback is not taking advantage of the symmetry. On the other hand, Routine 3 is used with CM reordering and experiments show it scales remarkably. Overlap is obtained over the time taken by the multiplication of the main diagonal, which requires the main diagonal to be stored separately as a preprocessing step. This work has the limitation of using very small test matrices and considering only symmetric matrices. Our solution achieves up to $19\times$ speedup compared to the sequential algorithm, where the matrix with the least amount of nonzeros we use has 18 times the nonzeros of the matrices used in \cite{ETH}'s  experiments, whereas, our largest one contains 84 times the nonzeros. We further develop the idea of reordering by splitting band form based on sparsity into three sparse and dense regions, and storing them separately in preprocessing to help speed up iterative solvers which use SS SpMVs. Splitting dense and sparse regions over a sparse matrix was studied earlier in SplitMV\cite{splitSPMV} , a parallel solver for sparse systems involving banded preconditioners which exhibit multi-level-parallelism, and improved speedups in 6- and 8-cores systems. Our parallel solution processes significantly more amount of communication and computation, which also shows we benefited more from overlapping. 

We frequently use  SPARSKit\cite{SPARSKIT}  software in our preprocessing stages when converting data storage formats is needed.
\begin{figure}[H]
\centering
  \includegraphics[width=0.9\columnwidth]{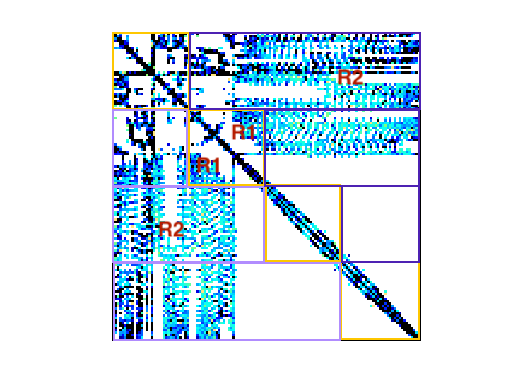}
  \caption{Matrix: Audikw$\_$1\protect\cite{SPARSKIT} . Assuming 4 parallel processes with block distribution, this is an illustration of our data decomposition. Conflicting regions are reflected with purple colored squares (R2). Yellow squares denote safe regions (R1) for the concurrency. Multiplying elements of R2 with the corresponding elements in input vector races on output vector with those of R2 pair that is located in the transpose region, as data are being written onto the same output location by different processes. Determining an element as conflicting is found by checking its column offset in the current process, where each process owns a row block (at upper or lower triangular region) .}~\label{fig:figure2}
\end{figure}

\section{3. Skew-SSpMV}

Essence of our fast implementation to Skew-SSpmV is the splits we have created after obtaining the band form. A regular block based data decomposition can predetermine the indices of racing elements before the iterative computations start. Having the band form first, we can apply this knowledge to our decomposition (splits) too. We exploit this fact by separately defining the conflicting regions in our middle, outer, and diagonal splits. We also apply row decomposition to input and output vectors by assigning contiguous chunks of memory locations to processes.

Two nonzero elements are symmetric pairs in the sense that while one value is negative, the other is positive, by definition of  skew-symmetry. Such conflicting elements cause threads to race one another, and they are likely to yield undefined multiplication results.

 As we account for only the lower or upper triangular part of the band matrix (with SSS form), each element in a square region is used to update its symmetric pair's corresonding output value as well. This is a common strategy to reduce memory bandwith utilization when processing symmetric matrices. Therefore, all main diagonal elements in the diagonal split is safe to concurrently execute by any  processes at any time, due to not writing onto other processes' output elements.

To generalize this approach, Figure \ref{fig:figure2} demonstrates that pair of elements in the yellow squares are safe to run in parallel due to their multiplication values being written onto separate locations that map only into their own processes.

On the other hand, multiplication values of elements contained by purple (either light or dark purple, based on chosen triangular side) rectangles are the source of data conflicts as they map onto rows across processes. After the processes multiply and write results onto its portion of output vector, it also computes for the transpose of the element and attempts to write onto non-local portions of output vector. 

 Being transformed into band matrix, matrix nonzeros are densified in and around yellow squares, which promises a better parallelism. As more elements are gathered around yellow square regions, and the less elements are located in purple rectangles, the scenario gives more performance benefits. 

Having discussed the impact of various data patterns in band matrix to performance, increasing parallel execution units also puts a limit on the overall performance because the depth of process scope shrinks. As explained earlier, the more processors are used, the less surface area the yellow rectangles have, i.e, decreasing the probability of elements being located in safe regions, thus, having more conflicting elements. 

One can observe no elements in process $0$ can create data conflicts with any other processes. In the following sections, when we communicate a conflicting element to its pair process, we will order those messages from the last process and sent in the direction of root, $0^{th}$, process, to resolve potential deadlocks during blocking communication operations.

\subsection{3.1.1 Serial SSpMV}
\begin{figure}[H]
\centering
  \includegraphics[width=0.9\columnwidth]{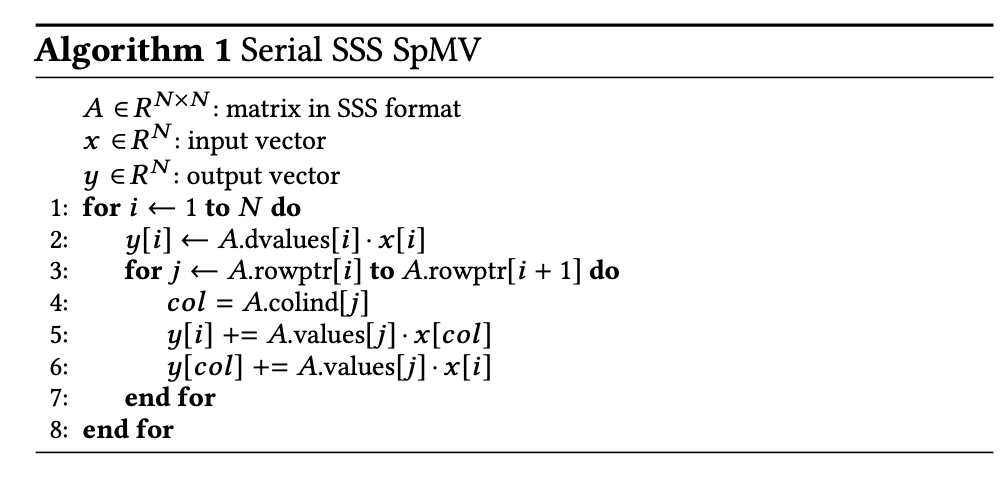}
  \caption{ Serial SSpMV with SSS }~\label{fig:figure3}
\end{figure}

The serial algorithm given in Figure \ref{fig:figure3} is adapted to skew-symmetry in our work, also referred to as our baseline solution. In this paper we refer to minimal iterations obtained by using SSS pointers as unrolling SSS data. It demonstrates the efficient unrolling of SSS data with $\Theta$(NNZ) time complexity to read elements and multiplies both pairs of the entries.

\subsection{3.1.2 Parallel 3-Way Banded Skew-SSpMV with RCM }
Having introduced the main cause of performance bottleck in the kernel, we initially process matrices in MATLAB into a banded form by using RCM algorithm with $\mathbf{\theta}$(NNZ) complexity. Then, we make use of SPARSKit library to transform band matrices into SSS storage format. Our preprocessor program then splits this band matrix by reading SSS format into 3 different data storage schemes. First split is the main diagonal, which we shall name as inner bandwidth of size $1$. Next, we have the relatively sparse part of the matrix, composed by the elements in the middle of band format, that we store excluding both the main diagonal and the outer elements in the remaining bandwidth. Outer elements compose the smallest split because they are scattered around the band form and mostly exist in intermediate processors. We have empirically picked outer bandwidth as 3 consecutive elements in the row-major order, i.e. outer bandwidth=3. We suggest that its size may be best determined by considering the total bandwidth and density characteristics of the generated band matrix. Our benchmark matrices are obtained from the Suite Sparse Matrix Collection \cite{suiteSparse}  and they are chosen to be the same set of matrices in \cite{Athens}  to correlate well with the provided performance results and to fix total number of conflicting operations. Table \ref{tab:tab1}. shares details of characteristics such as number of rows, number of nonzero elements, as well as the bandwidth after RCM reordering.

This part provides the specifics of our algorithm, Parallel 3-Way Banded Skew-SSpMV. Its inputs are the 3 divisions of the band matrix data in SSS format. Middle part, which contains the majority of matrix data, is distributed to MPI processes in blocked SSS format, which is analogous to slicing SSS for each processor. To discover conflicting elements, we first iterate over SSS data with $\mathbf{\theta}$(NNZ) time complexity to mark the conflicting process IDs paired with the current process ID based on the read column index. Given the conflicting process ID, corresponding portion of the X vector slice is communicated to the processes to be used in multiplications. We put an order on this part of the algorithm as processes are very likely to need immediate neighbour process' X portion to multiply with, due to RCM structure. This part is quickly done by letting last process P send its local X data to process P-1, and P-1 to P-2, and so on. Root will only receive  process 1's local X portion to prevent a deadlock. This is the second stage of communicating X data between processes and its communication cost tends to highly depend on the band structure. 

\begin{figure}[H]
\centering
  \includegraphics[width=0.7\columnwidth]{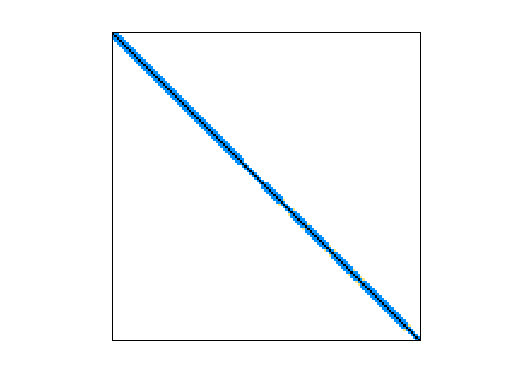}
  \caption{RCM-transformed Matrix: boneS10\protect\cite{SPARSKIT} }~\label{fig:figure4}
\end{figure}
\begin{figure}[H]
\centering
  \includegraphics[width=0.9\columnwidth]{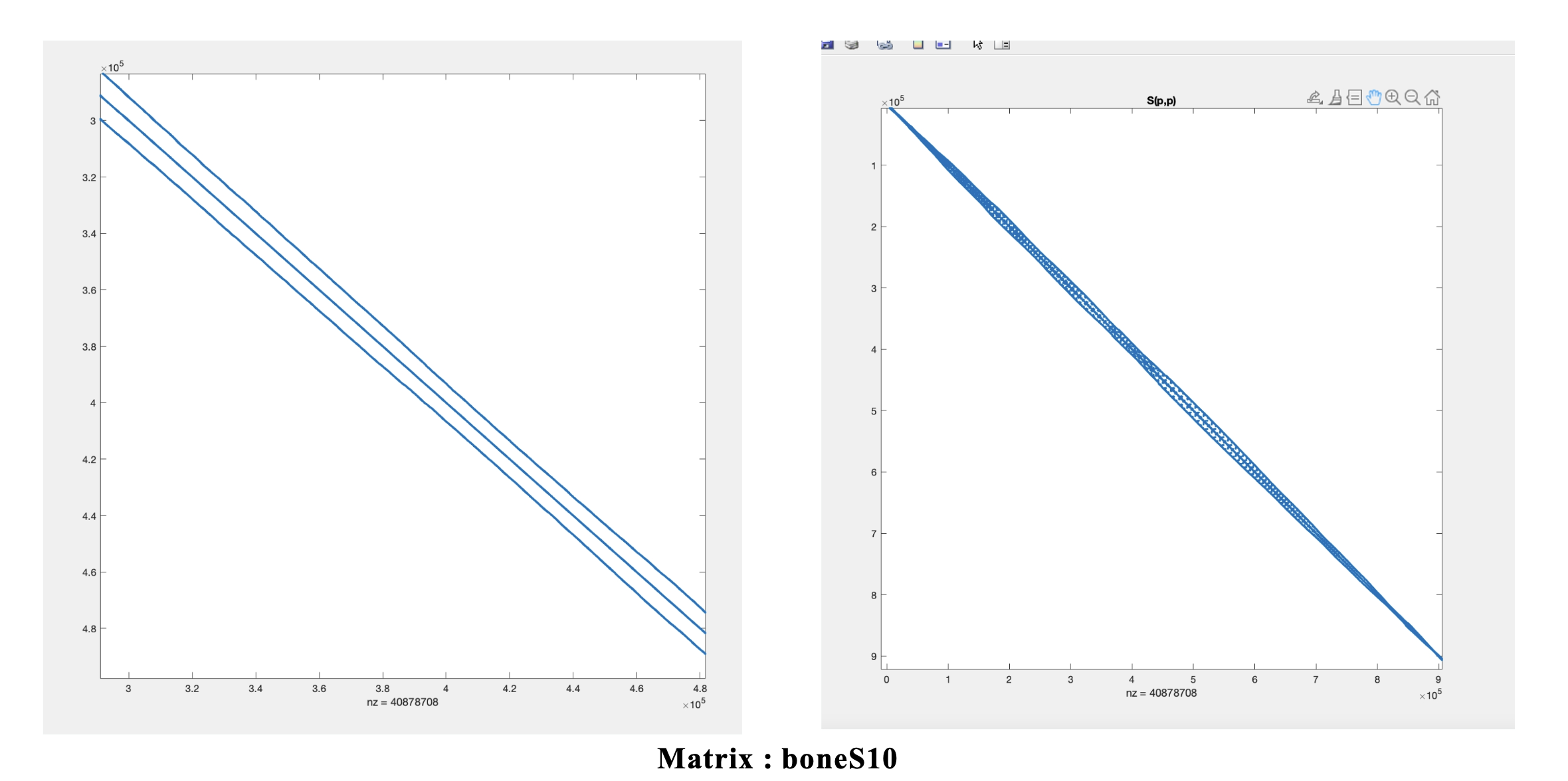}
  \caption{Effectiveness of RCM depends on the initial matrix structure in regards to reducing its bandwith and reorganizing the data in a more compact way. The less nonzeros a matrix has, the more trivial it is to restructure it to have reduced bandwith. }~\label{fig:figure5}
\end{figure}

\begin{figure}[H]
\centering
  \includegraphics[width=0.9\columnwidth]{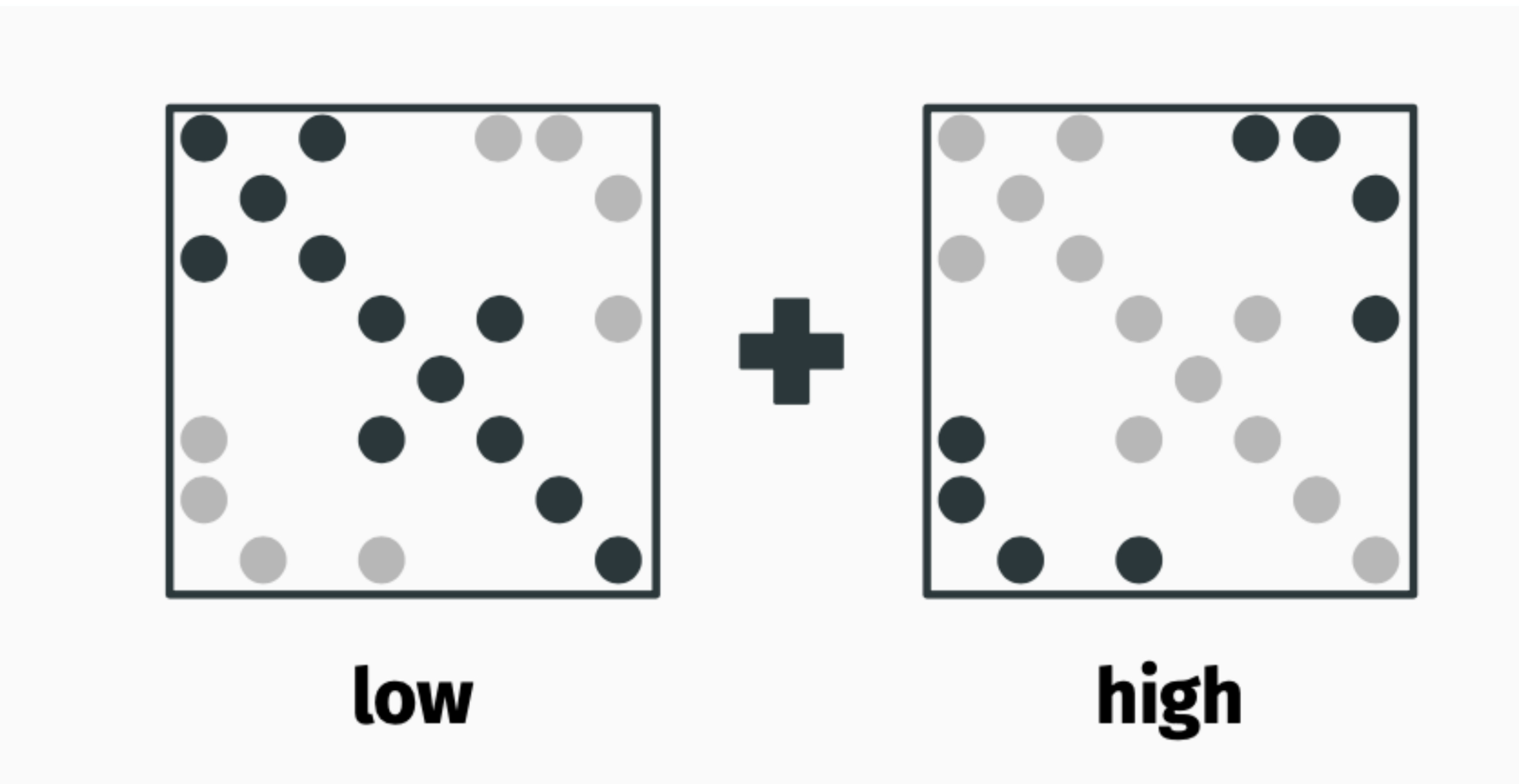}
  \caption{Bandwith split example over high and low bandwith elements in the same matrix to increase spatial locality when reading from cache. \protect\cite{Athens}}  ~\label{fig:figure6}
\end{figure}
\begin{figure}[H]
\centering
  \includegraphics[width=0.9\columnwidth]{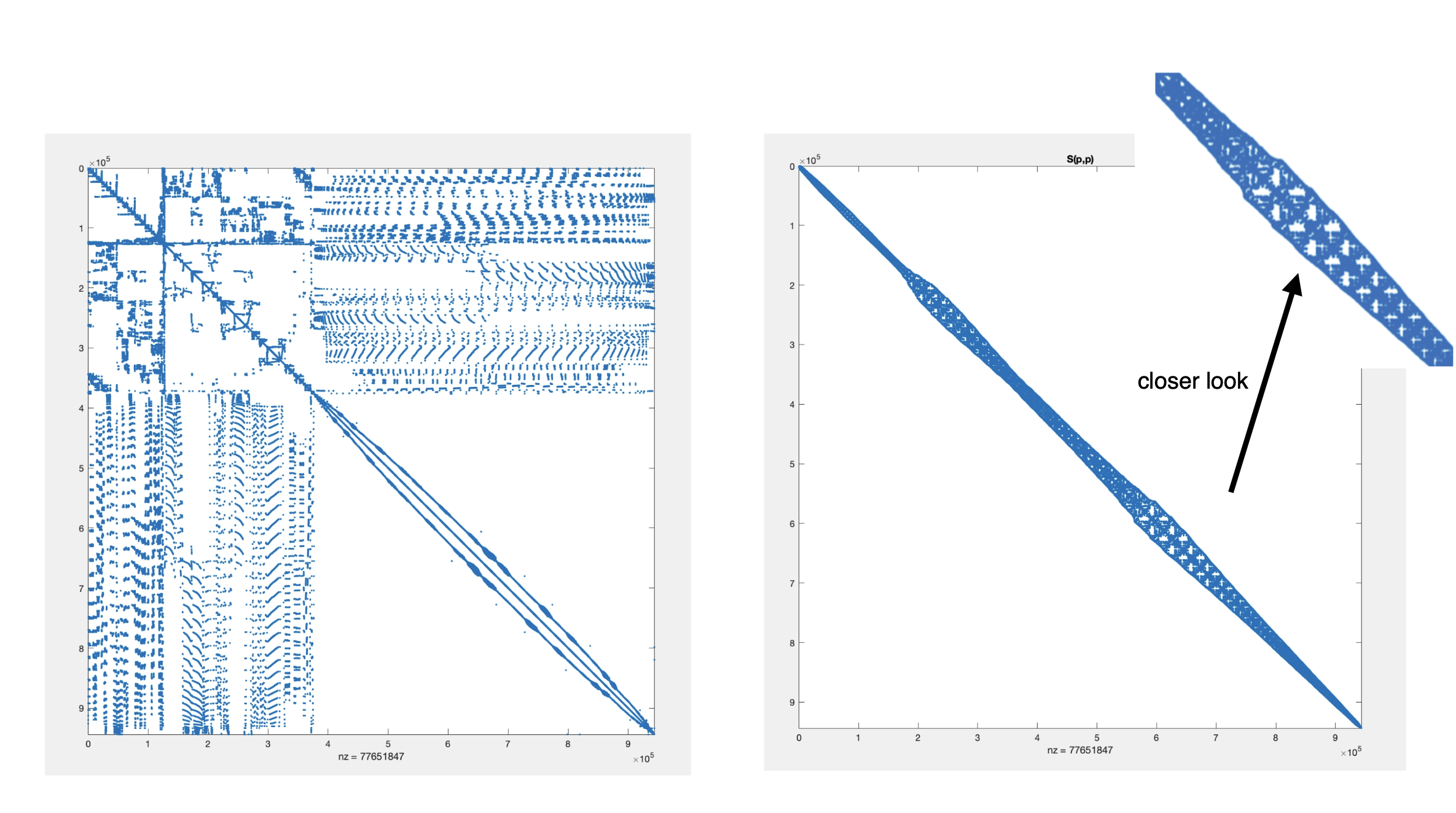}
  \caption{Closer look after RCM transformation, where the middle bandwith elements exhibit sparse placements, as opposed to diagonal and high bandwith elements, which are similarly dense in structure. We exploit bandwith splitting over band form with this observation.}  ~\label{fig:figure7}
\end{figure}
\begin{figure}[H]
\centering
  \includegraphics[width=0.9\columnwidth]{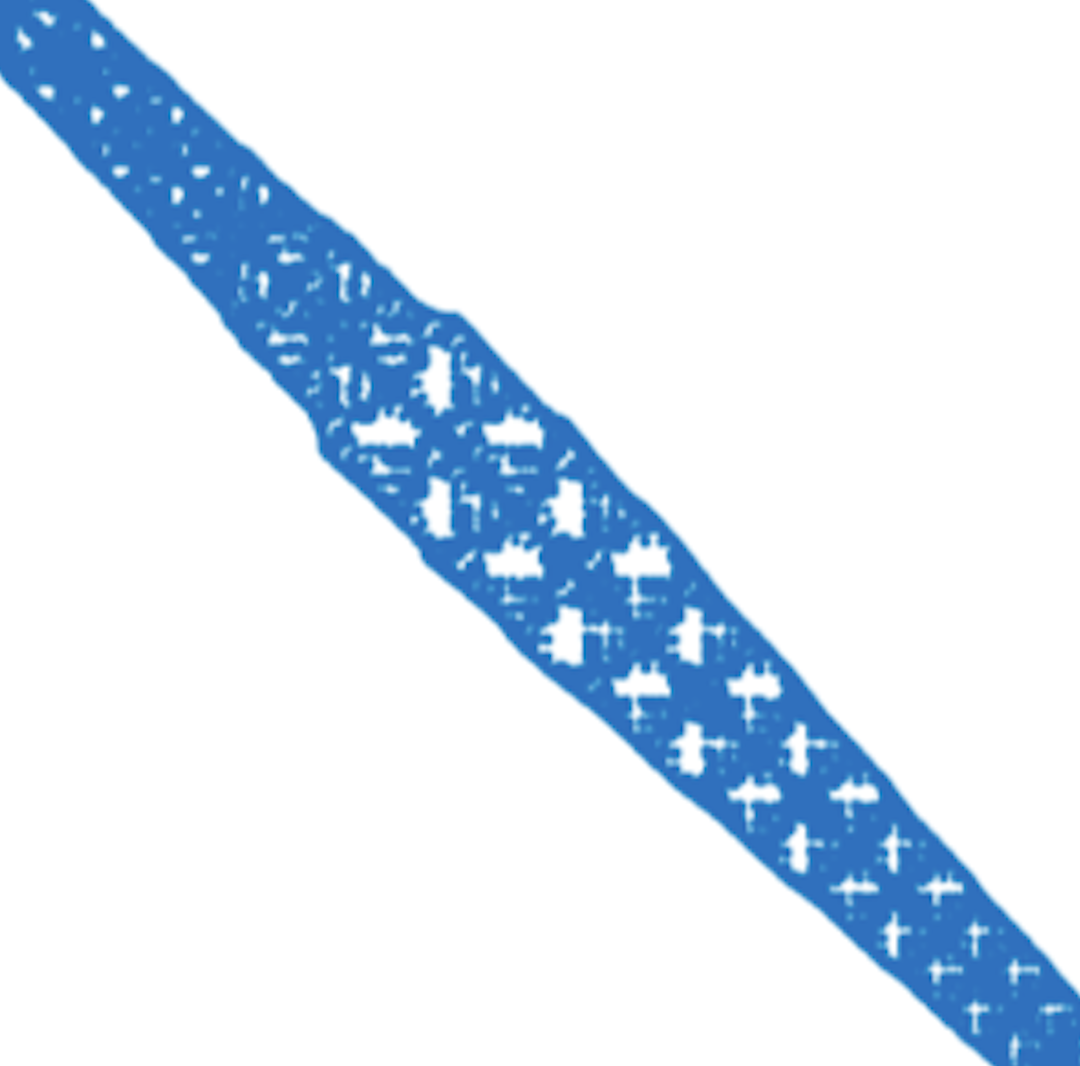}
  \caption{A closer look to general pattern of the band matrix Audikw$\_$1 after RCM transformation. We split either lower or upper half of such band structure into three parts. The main diagonal is the dense and smallest split. The middle part contains majority of the data in the entire matrix, but it also has a very sparse structure. Middle split mostly covers data in non-racing regions under our partitioning approach (See yellow squares in Figure \ref{fig:figure2}, explained in Part 3.) The distance from main diagonal to outer split is determined with a user specified bandwith. This is to provide flexibility to achievable performance in cases where the sparsity structure and band size of the band matrix is well predicted.}~\label{fig:figure8}
\end{figure}

There are more advantages in adopting such a scheme where we first detect conflicting items by checking its coordinates. Our major contribution to the parallel version of SSpMV algorithm by using SSS lies within the scheme described above. After communication part is complete, multiplication takes place. In our version of SSpMV algorithm with SSS, in each process, local middle split elements are iterated to be processed.  The key point is to reuse a numeric value in lower or upper triangle of matrix to be multiplied also with the symmetric pair's vector data. Once an element is read to be located in these purple rectangles, as demonstrated in Figure \ref{fig:figure2}, we also know about its mapping to the symmetric pair location's offset and process ID in output vector. For example, assuming we have a matrix of size 4x4, where each process owns a single row of matrix only in the lower triangular region, reading $\mathbf{A_{30}}$ with rowIndex=3 and columnIndex=0 assigns it to a conflicting rectangle. With one single read of this item by using SSS form, we also know about the value and exact coordinates of its symmetric pair. This also means 2 different multiplication can be computed and accumulated onto 2 different output locations in vector Y with a single item read. Below are the steps taken for this example. $\mathbf{A_{30}}$ is read once and $\mathbf{X_{0}}$ is communicated at the second stage of data movement for X vector. Note that $\mathbf{X_{3}}$ is locally owned by third process after the first stage of data movement is complete.
\begin{equation}
A_{30} = -A_{03}
\end{equation}
\begin{equation}
A_{30}*X_{0} = mul_1
\end{equation}
\begin{equation}
A_{03} *X_{3} = mul_2
\end{equation}
\begin{equation}
Y_{3} += mul_1
\end{equation}
\begin{equation}
Y_{0} += mul_2
\end{equation}
After the read element is known to be a conflicting one within the unsafe regions of its process, we locally store its multiplication results, $\mathbf{mul_2}$ , to be communicated to symmetric process which has ownership of $\mathbf{Y_{0}}$ and therefore can locally update $\mathbf{Y_{0}}$.  This part can be implemented with pair to pair communication as described, however we preferred to use one sided communication with MPI Accumulate due to its advantage of being an asynchronous operation. It is a Remote Memory Access routine and classified as non-blocking that can provide overlapping of communication with computation. Moving forward with the last part of the algorithm, for the processing of the final split of outer data, every process sequentially multiplies these negligible amounts of data. We rely on the benefits of sequential execution on CPUs for this last part only by considering how significantly small the total number of outer elements is and how closely located they are in memory. Outer part elements are most likely to be  conflicting elements that require further processing and therefore incurs abovementioned  communications costs in our proposed method when parallel processes. In sequential way, our approach avoids extra costs of fine grained arbitrary communication costs that would also incur irregular memory access penalties for these outer elements. Overall time complexity in our parallel solution is maintained at  $\mathbf{\theta(\frac{NNZ}{P})}$ where P represents number of MPI processes.

To balance the load among processes, we use block distribution to parallel execution units that scatters equal amount of rows in band matrix. Alternatively, one might consider distributing equal amount of non-zero elements to processes with unequal amount of rows, however, its benefits may not be as trivial to derive. Equal amount of nonzero elements does not necessarily mean sufficiently homogeneous distribution of conflicting and non-conflicting elements, therefore, corresponding communication and computation costs require careful examination which is tightly related to the spatial placements in band matrix form. We prefer to equally distribute rows with the reasoning of assigning preprocessing tasks to root eventually somewhat balances the termination time of execution units, as there is larger amount of work assigned to intermediate processes due to underlying band structure. We demonstrate block distribution approach achieves a sufficient parallel speedup in our platform. 

\section{4. Experiment Results}
We obtain our speedup plots by repeating our experiments more than few times to balance out execution times and taking their mean value. We also show the ideal speedup on the same plot. For parallel runs, we bind MPI processes to available 8 sockets in our platform with NUMA architecture. Our experiment platform contains 4 physical AMD Opteron CPU nodes with 16 cores in each (total of 64 cores). As we target iterative linear solvers and improving their performance, our results corresponds to the wall clock time that is taken by parallel multiplication algorithm only, which purposely excludes the time required for preprocessing operations of matrices since this overhead typically can be amortized in many repeated runs with the same matrix.

\subsection{4.1 Results of Parallel 3-Way Banded Skew-SSpMV}

\begin{figure}[H]
\centering
  \includegraphics[width=0.9\columnwidth]{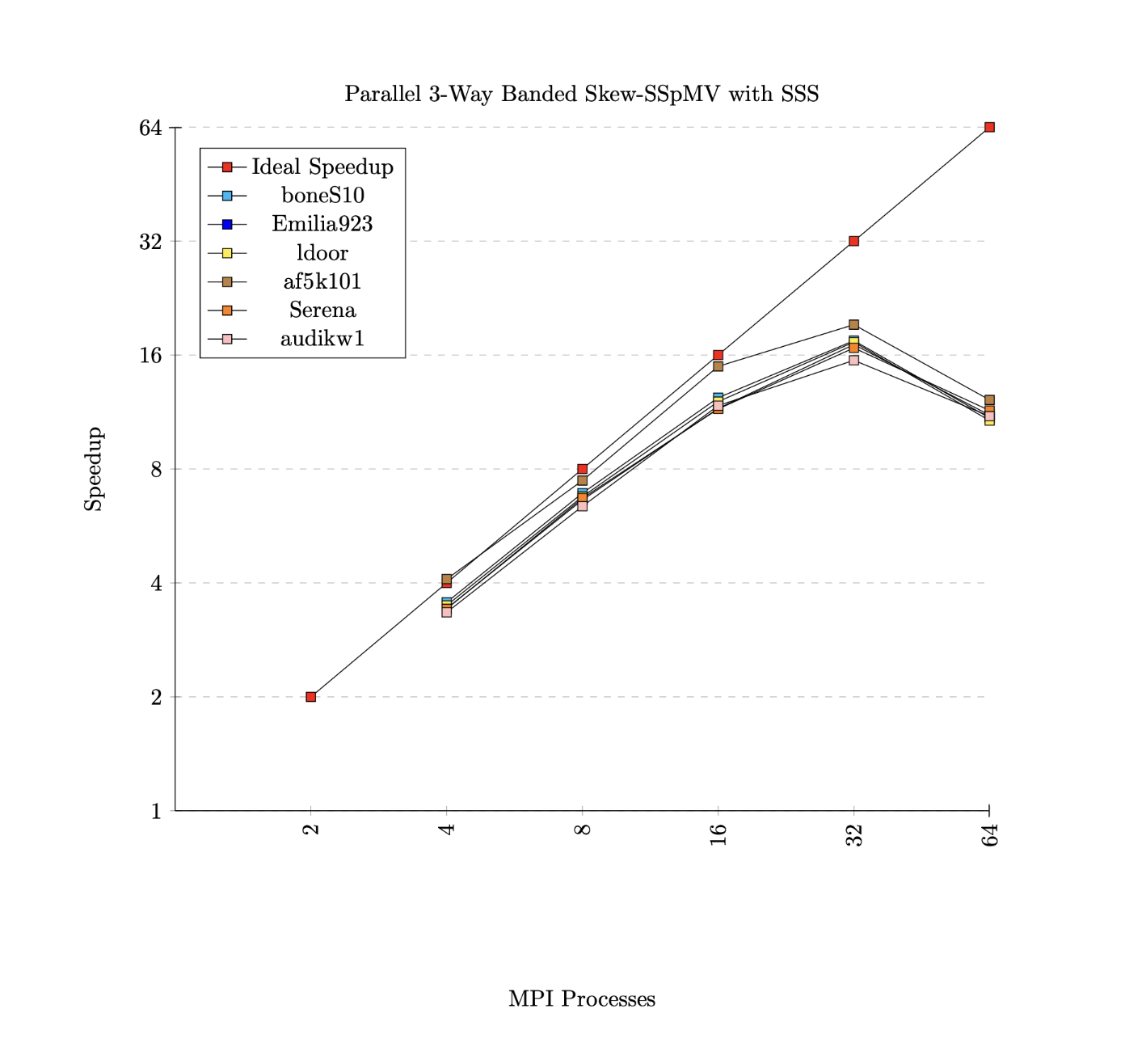}
  \caption{Speedup Results of our Solution}~\label{fig:figure9}
\end{figure}
Our implementation has strong scaling properties that are clearly better than using synchronization points, even with larger cores than reported in \cite{Athens}. For matrices Serena, af$\_$5$\_$k101, audikw$\_$1 and Serena, we have remarkably improved speedups.

Speedups we obtained also correlate well with the provided characteristics of the matrices in Table \ref{tab:tab1}. In particular, the best scaling at 19x is obtained for matrix af$\_$5$\_$k101 which contains the smallest number of nonzeros and the least RCM bandwith, as opposed to the poor scaling given in \cite{Athens} for this matrix. Serena and audikw$\_$1 are the heaviest matrices of the benchmark with the most amount of nonzeros and RCM bandwith. Their improvement tells us our method of partitioning, communicating and load balancing is effective when handling relatively bigger sparse matrices.

Experimenting with larger number of cores than \cite{Athens} is expected to introduce more communication overhead with the same input size. However our obtained improvements reflect that we handle these communication patterns efficiently, as well as 3-way partitioning and processing method being promising for iterative solvers.

For the rest of the matrices in benchmark, we remark that some matrices such as the one shown in Figure \ref{fig:figure5}, before preprocessing, their original structure is already similar to the band form that we usually obtain after RCM transformations. For such matrices, we expect to see less effect of such transformation. In this regard, a future work that can recognize and exploit original matrix patterns can be well integrated to our approach. 

As future work direction, a bandwith splitting storage mapping that could pre-identify data races at runtime given a number of processors could be implemented as a simple framework to abstractly support our multithreaded SkS-SpMV over different benchmarks, in the ways that we presented. With this abstract representation support, open-source linear algebra libraries such as OpenBLAS could support our work as a multithreaded routine for the adaptation of various iterative solvers.

\section{5. Conclusion}
In this paper, we propose a parallel implementation of skew-symmetric SpMVs in MPI and target increasing the performance of iterative linear solvers used by various scientific domains. Skew-symmetric matrices are easily transformable to symmetric matrices, however, their natural occurrences in algebra problems cannot be neglected. Our approach improves the performance of kernel with preprocessing of inputs by using RCM reordering and matrix splitting. We also emphasise preprocessing can be preferrable for iterative solvers more than incurring synchronization due to enormous conflicting elements in sparse matrices.  Although the performance improvement is highly dependent on the sparsity structure of the original matrix, reordering is quite effective to obtain a common form over many sparse matrices. Our solution offers an automated approach for many sparse matrices by using the bandwidth parameter so that the kernel can be best parallelized accordingly.

\section{6. Acknowledgements}
This work was mostly done when the first author was affiliated with Department of Computer Engineering, Middle East Technical University, Ankara, Turkey. \\
Department of Computer Science at University of Illinois has been officially renamed as the School of Computing and Data Science in 2024.

\bibliographystyle{SIGCHI-Reference-Format}
\bibliography{sample}


\begin{thebibliography}{00}


\ifx \showCODEN    \undefined \def \showCODEN     #1{\unskip}     \fi
\ifx \showDOI      \undefined \def \showDOI       #1{{\tt DOI:}\penalty0{#1}\ }
  \fi
\ifx \showISBNx    \undefined \def \showISBNx     #1{\unskip}     \fi
\ifx \showISBNxiii \undefined \def \showISBNxiii  #1{\unskip}     \fi
\ifx \showISSN     \undefined \def \showISSN      #1{\unskip}     \fi
\ifx \showLCCN     \undefined \def \showLCCN      #1{\unskip}     \fi
\ifx \shownote     \undefined \def \shownote      #1{#1}          \fi
\ifx \showarticletitle \undefined \def \showarticletitle #1{#1}   \fi
\ifx \showURL      \undefined \def \showURL       #1{#1}          \fi

\bibitem{splitSPMV}
{Eric Cox}, {Faisal Saied}, {Murat Manguoğlu}, {and} {Ahmed Sameh}. 2010.
\newblock A Split Sparse/Dense Matrix-vector Multiplication Routine.  (July
  2010).
\newblock
\newblock
\shownote{SIAM Annual Meeting Pittsburgh, Pennsylvania.}


\bibitem{suiteSparse}
{Timothy~A. Davis} {and} {Yifan Hu}. 2011.
\newblock \showarticletitle{{The University of Florida Sparse Matrix
  Collection}}.
\newblock {\em ACM Transactions on Mathematical Software Article 1\/}  {38}
  (2011), 25.
\newblock


\bibitem{Athens}
{Athena Elafrou}, {Georgios Goumas}, {and} {Nectarios Koziris}. 2019.
\newblock \showarticletitle{Conflict-free symmetric sparse matrix-vector
  multiplication on multicore architectures}. In {\em Proceedings of the
  International Conference for High Performance Computing, Networking, Storage
  and Analysis}. 1--15.
\newblock


\bibitem{ETH}
{Roman Geus} {and} {Stefan R{\"o}llin}. 2001.
\newblock \showarticletitle{Towards a fast parallel sparse symmetric
  matrix--vector multiplication}.
\newblock {\it Parallel Comput.} {27}, 7 (2001), 883--896.
\newblock


\bibitem{guducu2022non}
{Candan Güdücü}, {Jörg Liesen}, {Volker Mehrmann}, {and} {Daniel~B Szyld}.
  2022.
\newblock \showarticletitle{On non-Hermitian positive (semi) definite linear
  algebraic systems arising from dissipative Hamiltonian DAEs}.
\newblock {\em SIAM Journal on Scientific Computing\/} {44}, 4 (2022),
  A2871--A2894.
\newblock


\bibitem{idema2007minimal}
{Reijer Idema} {and} {Cornelis Vuik}. 2007.
\newblock {\em A minimal residual method for shifted skew-symmetric systems}.
\newblock Delft University of Technology.
\newblock


\bibitem{jiang2007algorithm}
{Erxiong Jiang}. 2007.
\newblock \showarticletitle{Algorithm for solving shifted skew-symmetric linear
  system}.
\newblock {\em Frontiers of Mathematics in China\/} {2}, 2 (2007), 227.
\newblock


\bibitem{irregular}
{Milind Kulkarni}, {Martin Burtscher}, {Rajeshkar Inkulu}, {Keshav Pingali},
  {and} {Calin Cas{\c{c}}aval}. 2009.
\newblock \showarticletitle{How much parallelism is there in irregular
  applications?}
\newblock {\em ACM sigplan notices\/} {44}, 4 (2009), 3--14.
\newblock


\bibitem{iterativeSolver}
{Volker Mehrmann} {and} {Murat Manguoğlu}. 2021.
\newblock \showarticletitle{{A two-level iterative scheme for general sparse
  linear systems based on approximate skew-symmetrizers. }}.
\newblock {\em Electronic Transactions on Numerical Analysis\/} (2021),
  370–391.
\newblock
\showDOI{%
\url{http://dx.doi.org/10.1553/etna_vol54s370}}


\bibitem{exahype}
{A. Reinarz}, {D.~E. Charrier}, {M. Bader}, {L. Bovard}, {M. Dumbser}, {K.
  Duru}, {F. Fambri}, {A.A. Gabriel}, {J.M. Gallard}, {S. Köppel}, {L. Krenz},
  {L. Rannabauer}, {L. Rezzolla}, {P. Samfass}, {M. Tavelli}, {and} {T.
  Weinzierl}. 2020.
\newblock \showarticletitle{ExaHyPE: An engine for parallel dynamically
  adaptive simulations of wave problems}.
\newblock {\em Computer Physics Communications\/} {254}, ISSN 0010-4655 (2020).
\newblock


\bibitem{SPARSKIT}
{Yousef Saad}. 1990.
\newblock {\em SPARSKIT: A basic tool kit for sparse matrix computations}.
\newblock {T}echnical {R}eport. Research Institute for Advanced Computer
  Science, NASA Ames Research Center.
\newblock


\end{thebibliography}


\end{document}